\documentclass[aps,prd,amsmath,amssymb,amsfonts,floats,floatfix,superscriptaddress,nofootinbib,notitlepage]{revtex4-1}
\usepackage{amsmath,dsfont,slashed,tensor,amssymb,epsf,epsfig,amsthm,bm,graphicx}
\usepackage{hyperref}
\usepackage{color}
\definecolor{darkgreen}{rgb}{0,0.35,0}

\begin{document}
\global\parskip 6pt

\author{Alfredo Guevara}
\email{alfguevara@udec.cl}
\affiliation{Centro de Estudios Cient\'{\i}ficos (CECs), Av. Arturo Prat 514, Valdivia, Chile}
\affiliation{Departamento de F\'isica, Universidad de Concepci\'on, Casilla 160-C, Concepci\'on, Chile}

\author{Pablo Pais}
\email{pais@cecs.cl}
\affiliation{Centro de Estudios Cient\'{\i}ficos (CECs), Av. Arturo Prat 514, Valdivia, Chile}
\affiliation{Physique Th\'eorique et Math\'ematique,
Universit\'e Libre de Bruxelles and International Solvay Institutes,
Campus Plaine C.P. 231, B-1050 Bruxelles, Belgium}

\author{Jorge Zanelli}
\email{z@cecs.cl}
\affiliation{Centro de Estudios Cient\'{\i}ficos (CECs), Av. Arturo Prat 514, Valdivia, Chile}

\title{Dynamical Contents of Unconventional Supersymmetry}

\begin{abstract}
The Dirac Hamiltonian formalism is applied to a system in $(2+1)$-dimensions consisting of a Dirac field $\psi$ minimally coupled to Chern-Simons $U(1)$ and $SO(2,1)$ connections, $A$ and $\omega$, respectively. This theory is connected to a supersymmetric Chern-Simons form in which the gravitino has been projected out (unconventional supersymmetry) and, in the case of a flat background, corresponds to the low energy limit of graphene. The separation between first-class and second-class constraints is performed explicitly, and both the field equations and gauge symmetries of the Lagrangian formalism are fully recovered. The degrees of freedom of the theory in generic sectors shows that the propagating states correspond to fermionic modes in the background determined by the geometry of the graphene sheet and the nondynamical electromagnetic field. This is shown for the following canonical sectors: i) a conformally invariant generic description where the spinor field and the dreibein are locally rescaled; ii) a specific configuration for the Dirac fermion consistent with its spin, where Weyl symmetry is exchanged by time reparametrizations;  iii) the vacuum sector $\psi=0$, which is of interest for perturbation theory. For the latter the analysis is adapted to the case of manifolds with boundary, and the corresponding Dirac brackets together with the centrally extended charge algebra are found. Finally, the $SU(2)$ generalization of the gauge group is briefly treated, yielding analogous conclusions for the degrees of freedom.
\end{abstract}

\date{\today}
\maketitle

\section{Introduction}       
\label{introduction}

Supersymmetry (\textbf{SUSY}) is a natural --and perhaps unique-- way to unify internal and spacetime symmetries in the description of fundamental particles and interactions. In spite of its elegant appeal, it is puzzling that no evidence of supersymmetry has been seen in the current phenomenology.  In the seminal work of Wess and Zumino, an action principle based on the idea of a supergauge symmetry was examined in a Lagrangian consisting of spin-0 and spin-1/2 fields, the Wess-Zumino  (\textbf{WZ}) model. The conclusion there was that in order for this symmetry to close, its parameter had to be a Killing spinor of the background spacetime \cite{WZ}. This indicates that the existence of supersymmetry requires spacetime itself to possess some symmetry that allows for the existence of some sort of Killing spinors. Indeed, the superalgebra behind the WZ model is an extension of the Poincar\'e algebra, whose geometric interpretation calls for a maximally symmetric spacetime, namely Minkowski space.

In the WZ model and in most supersymmetric particle models, the fields form an irreducible vector representation of the super-Poincar\'e algebra, a supermultiplet. This implies that bosons and fermions come in pairs with equal mass and other quantum numbers, but with spins differing by $\hbar/2$ (superpartners). Since no such duplication of the particle spectrum has been observed, it is argued that SUSY must be broken at the energy scales that we have been able to explore, but it is supposedly restored at a sufficiently high energies. An alternative to this picture, where the fields do not form a vector multiplet but rather enter as parts of a connection can also be considered, and in that case degenerate superpartners are not necessarily  present \cite{AVZ,APZ}. This is a generic feature, for example, of supersymmetric Chern-Simons (\textbf{CS}) theories, where bosonic and fermionic fields combine to form a connection for the supersymmetric graded Lie algebra \cite{Ch,BTrZ,TZ}.

It is well known that CS theories in three dimensions for any Lie algebra have no local degrees of freedom \cite{BGH}.  This is true also for CS theories based on graded Lie algebras \cite{Chandia}, like in the case of the CS supergravity for the $\mathfrak{osp}(2|2)$ algebra. By contrast, a massive spin-1/2 field in a fixed three-dimensional background of has $2n$ propagating degrees of freedom, where $n=1$ for Majorana and $n=2$ for Dirac spinors \cite{HRT,HT}. Now, if in the $\mathfrak{osp}(2|2)$ CS theory the gravitino field $\chi^\alpha_\mu$ is split into a spin-$1/2$ Dirac spinor $\psi^\alpha$ and the vielbein $e^a_\mu$, the fermionic sector of the reduced theory describes a Dirac fermion in a curved background, minimally coupled to $\mathfrak{u}(1)$ and $\mathfrak{so}(2,1)$ gauge connection one-forms $A=A_\mu dx^\mu$ and $\omega^a{}_b=\omega^a{}_{b\mu} dx^\mu$, respectively \cite{AVZ}. It is therefore only natural to inquire whether this reduced theory has zero local degrees of freedom (\textbf{DOF}) as the original CS system, or has four local degrees of freedom of a spin-$1/2$ Dirac fermion. The question is further complicated by the fact that in the reduced Lagrangian the dreibein are not Lagrange multipliers (their time derivatives $\dot{e}^a$ appear explicitly in the action) and therefore $e^a_\mu$ are in principle dynamical fields as well.

The identification of the physical degrees of freedom can be addressed by direct application of Dirac's analysis of constrained Hamiltonian systems \cite{Dirac}, which systematically separates the dynamical fields from the gauge degrees of freedom. In the case of CS theories, however, the separation between first and second-class constraints is a delicate issue, and the system considered here is not an exemption. The reduced action in \cite{AVZ} reads
\begin{equation}\label{I}
I[\psi,e,A,\omega]= \int \frac{1}{2}\Bigl[ \overline{\psi}\slashed{e} (\overleftarrow{D} - \overrightarrow{D})\slashed{e} \psi + AdA + \frac{1}{2} \omega^a{}_b d \omega^b{}_a + \frac{1}{3} \omega^a{}_b \omega^b_c \omega^c_a \Bigr]\, ,
\end{equation}
where $\slashed{e} \equiv \Gamma_a e^a = \Gamma_a e^a_\mu dx^\mu$ and $e^a_\mu$ are the dreibein (see Appendix \ref{conventions} for notation). In addition to the local $U(1)\times SO(2,1)$ symmetry and spacetime diffeomorphisms, this action is invariant under local Weyl rescalings,
\begin{equation}\label{Weylrscl}
e^a_\mu \rightarrow \lambda e^a_\mu \,, \quad \psi \rightarrow \lambda^{-1} \psi \,, \quad \bar{\psi} \rightarrow \lambda^{-1}\bar{\psi} \,,
\end{equation}
where $\lambda(x)$ is a non-vanishing, real and differentiable function. All of these symmetries are in principle associated with first-class constraints that reduce the physical phase space.

Varying the action with respect to $\psi$ yields the Dirac equation with a mass term $m=\frac{1}{2|e|} \epsilon^{\mu\nu\rho} \eta_{ab}e^{a}_{\mu} D_\nu e^b_\rho$ (including hermiticity corrections), while varying with respect to $e^a_{\mu}$ implies the vanishing of the energy-momentum tensor, $\mathcal{T}^{\mu \nu}=\frac{1}{2|e|}\eta^{ab} E_a^{\mu}\frac{\delta L}{\delta e_{\nu}^{b}}+(\mu \leftrightarrow \nu)$, with $E_a^\mu$ the inverse dreibein. In particular, the vanishing of the trace $\mathcal{T}^{\mu}{}_\mu=0$ is consistent with the local scale invariance of the action.

For a fixed background the local degrees of freedom should correspond to the $2n$ independent components of the Dirac field in flat spacetime. A quick analysis suggests that six out of the nine components of the dreibein can be eliminated by the conditions $\mathcal{T}^{\mu\nu}=0$, while the remaining three can be gauged away via two spatial diffeomorphisms and a Weyl scaling. In CS theories, time diffeomorphisms are not independent, which means their phase space generators are linear combinations of the remaining first-class constraints \cite{BGH}.

As noted in \cite{AVZ}, the closure the supersymmetry for (\ref{I}) requires the parameter of the SUSY transformation to satisfy a subsidiary condition to ensure the variation $\delta \psi$ to have spin-1/2, like $\psi$ itself. This subsidiary condition is satisfied if the SUSY parameter is required to be a Killing spinor of the background and, like in the original WZ system, this means that supersymmetry is a global (rigid) symmetry \cite{APZ}. Since this is not a gauge symmetry, it is not generated by a first-class constraint that would further reduce the number of physical degrees of freedom.

\section{Hamiltonian analysis}        
Splitting the fields and their derivatives into time ($t$) and spatial components ($i,j=1,2$), the Lagrangian (\ref{I}) can be written, up to a boundary term, as
\begin{equation}
L  =  \epsilon^{ij}\Bigl[-\eta_{ab} \dot{e}_i ^a e_j^b \bar{\psi}\psi-\dot{\bar{\psi}}\Gamma_{ij}\psi+\bar{\psi}\Gamma_{ij}\dot{\psi} +\frac{1}{2}\eta_{ab}\dot{\omega}_i^a \omega_j ^b  + \frac{1}{2}\dot{A}_i A_j \Bigr] -e_t ^a K_a + \omega_t ^a J_a +A_t K\,,\label{eq:Splitlag}
\end{equation}
where we defined $\Gamma_{ij}:=e_{i}^{a}e_{j}^{b}\Gamma_{ab}$, $\omega^a := \frac{1}{2} \epsilon^{abc}\omega_{bc}$, and
\begin{eqnarray}
K_a & := & 2\epsilon^{ij}\Bigl[ \eta_{ab}T_{ij}^b \bar{\psi}\psi -e_i^b (\bar{\psi}\Gamma_a \Gamma_b \overrightarrow{D}_j \psi + \bar{\psi} \overleftarrow{D}_j \Gamma_b \Gamma_a \psi)\Bigr]\,,\label{Ka} \\
J_{a} & := & \epsilon^{ij}\eta_{ab}(\frac{1}{2}R_{ij}^{b}-\epsilon_{\:cd}^{b}e_{i}^{c}e_{j}^{d}\bar{\psi}\psi)\,, \label{Ja} \\
K & := & \epsilon^{ij}(\partial_{i}A_{j}-i\bar{\psi}\Gamma_{ij}\psi)\,, \label{K}
\end{eqnarray}
where $R^a= d\omega ^ a+ \epsilon^a_{\,\,\, bc}\omega^b \omega ^c$ is the Lorentz curvature and $T^{a}=de^{a}+\epsilon^a{}_{bc}\omega^b e^c$ is the torsion tensor of the background. Lagrangian (\ref{eq:Splitlag}) describes the evolution of $(21+4n)$ coordinate fields: $e^a_\mu$ (nine), $\omega^a_\mu$ (nine), $A_\mu$ (three), $\psi$ (2n) and $\bar{\psi}$ (2n); among them there are seven ($e^a_t$, $\omega^a_t$ and $A_t$), whose time derivatives do not appear in the Lagrangian and are therefore Lagrange multipliers with vanishing canonical momenta. For the remaining components the Lagrangian contains only first time derivatives and therefore each momentum is a function of the coordinate fields. Thus, the following $(14+4n)$ primary constraints are obtained
\begin{eqnarray}\label{primary_constraints}
\varphi_{a}^{i} & = & p_{a}^{i}+2\epsilon^{ij}\eta_{ab}e_{j}^{b}\bar{\psi}\psi\approx0\,,\nonumber \\
\Omega & = & \chi+\epsilon^{ij}\Gamma_{ij}\psi\approx0\,,\nonumber \\
\bar{\Omega} & = & \bar{\chi}-\epsilon^{ij}\bar{\psi}\Gamma_{ij}\approx0\,,\\
\phi_{a}^{i} & = & \pi_{a}^{i}-\frac{1}{2}\epsilon^{ij}\eta_{ab}\omega_{j}^{b}\approx0\,, \nonumber \\
\phi^{i} & = & \pi^{i}-\frac{1}{2}\epsilon^{ij}A_{j}\approx0\, .\nonumber
\end{eqnarray}

The seven combinations $K_a$, $J_a$, $K$ in (\ref{eq:Splitlag}) are then secondary constraints associated to the Lagrange multipliers. Moreover, the canonical Hamiltonian weakly vanishes and the total Hamiltonian can be taken as an arbitrary linear combination of all the constraints\footnote{Hereafter we perform the integrations over the spatial slices $\Sigma$ given by $t=constant$, for which we do not consider a boundary. The cases where $ \partial\Sigma \neq \emptyset$, which yield asymptotic charges, are discussed in Section \ref{bosonic_vacuum}.},
\begin{equation} \label{hamiltonian}
H_T = \int d^2 x\left[ e^a_t K_a -\omega^a_t J_a - A_t K  + \varphi^i_a \lambda^a_i +\phi^i_a \Lambda^a_i +\bar{\Lambda}\Omega + \bar{\Omega}\Lambda +\lambda_i \phi^i \right] \,.
\end{equation}

It can be proved that the following seven linear combinations are first-class constraints (see Appendix \ref{poisson_brackets} for details)
\begin{eqnarray}\label{first_class_constraints}
\tilde{J}_a & := & J_a +\epsilon^{}_{ac} \varphi^j_b e^c_j +\frac{1}{2}(\bar{\Omega}\Gamma_a \psi - \bar{\psi}\Gamma_a \Omega) + D_j \phi^j_a \,, \nonumber \\
\tilde{K} & := & K-\frac{i}{2}(\bar{\Omega}\psi-\bar{\psi}\Omega)+\partial_j \phi^j \,,\nonumber \\
\Upsilon & := & -e^b_j \varphi^j_b +\bar{\Omega}\psi+\bar{\psi}\Omega \,, \\
\mathcal{H}_i & := & e^a_i K_a -e^a_i D_j \varphi^j_a + T^a_{ij} \varphi^j_a + \bar{\psi} \overleftarrow{D}_i \Omega +\bar{\Omega}\overrightarrow{D}_{i}\psi-\omega_{i}^{a}\tilde{J}_{a}-A_{i}\tilde{K}+\phi^{j}F_{ij} +\phi_{a}^{j}R^a_{ij} \, . \nonumber
\end{eqnarray}
Here the (spatial) covariant derivative $D_i$ acts on each field according to its transformation properties, as in (\ref{def_covariant_derivative}). Using \eqref{Ka}-\eqref{primary_constraints}, the generators $\mathcal{H}_i$ can be expressed as
\begin{eqnarray*}
\mathcal{H}_i & = & \left(\partial_i A_j -\partial_j A_i \right)\pi^j -A_i \partial_j \pi^j + \left(\partial_i \omega^a_j -\partial_j\omega^a_i \right)\pi^j_a -\omega^a_i \partial_j \pi^j_a \\
 &  & +\left(\partial_i e^a_j -\partial_j e^a_i \right)p^j_a -e^a_i \partial_j p^j_a + \partial_i \overline{\psi} \chi + \overline{\chi}\partial_i \psi\,,
\end{eqnarray*}
which can be readily seen to generate spatial diffeomorphisms on phase space functions $F$ as $\{F, \int \xi^i \mathcal{H}_i  \} = \mathcal{L}_{\xi}F$. This in turn means that
\begin{equation}\label{diffeo_algebra}
\left\{\mathcal{H}_{i}(x), \mathcal{H}_{j}(y)\right\} = \mathcal{H}_{i}(y)\partial_{j}^{(y)}\delta^{(2)}(x-y)-\mathcal{H}_{j}(x)\partial_{i}^{(x)}\delta^{(2)}(x-y)\,,
\end{equation}
as expected from generators of spatial diffeomorphisms \cite{NT}. On the other hand, it can be directly checked that $\tilde{J}_a$, $\tilde{K}$ and $\Upsilon$ generate $SO(2,1) \times U(1) \times Weyl$ transformations over all the fields and momenta. Indeed, they satisfy the (weakly vanishing) Poisson relations (\ref{sec-prim}) with all the constraints, and one finds
\begin{eqnarray} \label{simplealg}
\{\tilde{J}_{a},\tilde{J}_{b}\} & = & \epsilon_{ab}^{\quad c}\tilde{J}_{c}\,,\nonumber \\
\{\tilde{K},\tilde{K}\} & = &\{\Upsilon,\Upsilon\} =\{\tilde{K},\Upsilon\} = 0\,, \\
\{\tilde{J}_{b},\tilde{K} \} &=& \{\tilde{J}_{b},\Upsilon\} = 0\,. \nonumber
\end{eqnarray}
Together with the generators of spatial diffeomorphisms these then form a first-class Poisson algebra.

Note that performing a shift in the Lagrange multipliers of the form
\begin{eqnarray}
\lambda^a_i \rightarrow \lambda'^a_i &=& -ve^a_i + \lambda^a_i \,,\nonumber \\
\Lambda'^\alpha \rightarrow  \Lambda'^\alpha &=& v\psi^\alpha + \Lambda^\alpha \,, \label{shift} \\
\overline{\Lambda}_\alpha \rightarrow  \overline{\Lambda}'_\alpha &=& v\overline{\psi}_\alpha + \overline{\Lambda}_\alpha \, , \nonumber
\end{eqnarray}
produces a shift in the total Hamiltonian \eqref{hamiltonian},
\begin{equation}
H_T \rightarrow H'_T = H_T + \int v \Upsilon d^2 x\, . \label{Hshift}
\end{equation}
This accounts for the Weyl invariance \eqref{Weylrscl} of the system. However, the absence of spatial derivatives in $\Upsilon$ implies that such symmetry is generated by a purely local constraint with no associated asymptotic charges, as explained in detail below (recent examples of this fact can be found in \cite{Jackiw-Pi} and references therein,  see \cite{IorioWeyl} for a thorough discussion). Weyl symmetry is thus a local redefinition of the fields without any observable effects. The corresponding symmetry breaking, however, leads to physical consequences as we will discuss.

\subsection{Generic conformally invariant sector}\label{generic_sector}  
We now assume that in a generic\footnote{Following \cite{BGH} we understand by \textit{generic} sectors those with a maximum number of degrees of freedom or, equivalently, a minimum number of independent first-class constraints.} background the $(14+4n)$ time preservation equations of the primary constraints fix an equal number of Lagrange multipliers (see Appendix \ref{solving_consistency} for details). The other seven parameters remain free in the total Hamiltonian \eqref{hamiltonian}, to form a linear combination of the first-class constraints. Choosing \{$\mathcal{H}_i$, $\tilde{J}_a$, $\tilde{K}$, $\Upsilon$\} as the basis of these generators, the total Hamiltonian can be written as
\begin{equation}\label{Ht1}
H_T = \int d^{2}x \left[\xi^{i}\mathcal{H}_{i} + v \Upsilon - \omega_t^{a}\tilde{J}_{a}-A_t \tilde{K} \right] \,,
\end{equation}
Here the Lagrange multipliers $\xi^i$, $v$, $\omega_t^a$ and $A_t$ are real and arbitrary functions on equal footing. As the Hamiltonian is a combination of first-class constraints, the time preservation relations are fulfilled by construction, and no additional (tertiary) constraints are produced in the Dirac algorithm. Note further that for any phase space function $F$ the Poisson bracket $\{F,H_T\}$ coincides with the corresponding Dirac bracket.

Now, the expression (\ref{Ht1}) was obtained from (\ref{hamiltonian}) by choosing
\begin{equation}\label{ea=xiei}
e^a_t = \xi^i e^a_i \, .
\end{equation}
This means that the three components $e^a_t$ are functions of the two free parameters $\xi^i $, while it also implies a degenerate dreibein, $|e|=0$. Although this may seem puzzling for a metric interpretation, it is dynamically consistent and allows to do the correct counting of the local degrees of freedom (see e.g., \cite{BGH} and Appendix \ref{solving_consistency}). The choice (\ref{ea=xiei}) is equivalent to the gauge $N^{\perp}=0$ in gravitation, which is perfectly acceptable as well as generic choices in ordinary gauge systems, i.e. Yang-Mills \cite{HTV, T}. Furthermore, it also allows to write the  generator of temporal diffeomorphisms as a linear combination of generators of local spatial diffeomorphisms, rescalings, Lorentz and $U(1)$ transformations\footnote{It can be explicitly shown that $\{\cdots,\int N\mathcal{H} \} \approx\mathcal{L}_{N\frac{\partial}{\partial t}}(\cdots) $, which is a general property of coordinate invariant systems \cite{HRT}.},
\begin{equation}
\mathcal{H}=\xi^i \mathcal{H}_i +v\Upsilon-\omega^a_t \tilde{J}_a -A_t \tilde{K} \, .
\end{equation}

Note that the degenerate condition $|e|=0$ remains invariant under local Weyl symmetry. Next, we consider a choice in which the Weyl symmetry is broken and the $e^a_t$ remains arbitrary so that the vielbein need not to be degenerate.

\subsection{Pure spin-1/2 generic sector}\label{half_spin_sector}      
We now examine a specific sector of the theory in which (\ref{ea=xiei}) is not imposed but the Weyl invariance is fixed instead. We consider a generic sector for the fields $e$ and $\psi$ that restricts the fermionic excitations to have spin-1/2 only. A fermionic field $\chi^\alpha_a$ transforms as a vector in the index $a$ and as a spinor in the index $\alpha$ and therefore belongs to a representation $1 \otimes 1/2=3/2 \oplus 1/2$ of the Lorentz group. There is a unique decomposition of this field into irreducible representations  $\chi_a=\chi^{(3/2)}_a + \chi^{(1/2)}_a$, where
\begin{eqnarray}
[ \delta^b_a - \frac{1}{3}\Gamma_a \Gamma^b ] \chi^{(1/2)}_b =0\, ,  \label{P3/2} \\
\Gamma^a   \chi^{(3/2)}_a =0\, .  \label{P1/2}
\end{eqnarray}

In the case of the field $\psi$, the condition that it only carries spin-1/2 requires that $D_{\mu}\psi$ also belongs to the spin-$1/2$ representation and should therefore be in the kernel of the spin-$3/2$ projector, namely,\footnote{Formally, if the scale has not been fixed the sector should be defined as the equivalence class of configurations satisfying \eqref{Proj-1/2} up to Weyl transformations. A manifestly covariant condition can be attained by introducing a gauge field for scale invariance $D_\mu \rightarrow D_\mu +W_{\mu}$, as originally proposed by Weyl \cite{IorioWeyl}.}
 \begin{equation} \label{Proj-1/2}
 [\delta^\mu_\nu-\frac{1}{3}\Gamma_\nu \Gamma^\mu] D_\mu \psi =0 \, ,
 \end{equation}
where $\Gamma_\mu\equiv e^a_\mu \Gamma_a$. This implies that the system does not generate local spin-$3/2$ excitations --no gravitini-- through parallel transport of the fermion. It may be regarded as a consistency condition for the system (\ref{I}) if it is meant to describe a Dirac spinor. The general solution of (\ref{Proj-1/2}) is
\begin{equation}
D_\mu \psi =\Gamma_\mu \xi\,  , \label{D-psi}
\end{equation}
where $\xi$ is an arbitrary Dirac spinor.

Next, in order to study the dynamical content of the sector, we perform a partial gauge fixing. As shown in \cite{AVZ}, the field equations for the action (\ref{I}) require the torsion to be covariantly constant, $DT^a=0$, where $D$ is the Lorentz covariant exterior derivative (see Appendix \ref{conventions}). The general solution of this equation, with an appropriate local rescaling of the dreiben --using the freedom due to Weyl symmetry-- is of the form
\begin{equation}
T^a = \alpha \epsilon^a{}_{bc}e^b e^c \,, \label{T=ee}
\end{equation}
where $\alpha$ is an arbitrary (dimensionful) constant. Now, inserting (\ref{D-psi}), (\ref{T=ee}) in \eqref{Ka} we obtain
\begin{equation}
K_a = 2\epsilon^{ij}e^b_i e^c_j \Bigl[2\alpha \epsilon_{abc} \bar{\psi}\psi - (\bar{\xi} \Gamma_a \Gamma_b \Gamma_c \psi- \bar{\psi}\Gamma_c \Gamma_b \Gamma_a \xi) \Bigr]\, . \label{Ka-psi}
\end{equation}
In order for the constraint condition $K_a \approx 0$ not to introduce additional restrictions on the fields, the right-hand-side of (\ref{Ka-psi}) must identically vanish. This demands $\xi=\alpha \psi$ and therefore, this selects the sector \footnote{The projector \eqref{Proj-1/2} is a generalization of the so-called `twistor operator', which defines conformal Killing spinors \eqref{D-psi} in the absence of torsion \cite{WZ,Baum}. Equation \eqref{Dpsi=apsi} can be regarded as the  Killing spinor equation for a curved background \cite{SU2}. Remarkably, \eqref{D-psi} and \eqref{Dpsi=apsi} are completely equivalent by virtue of the Dirac equation.}
\begin{equation}
D_{\mu}\psi = \alpha\Gamma_{\mu}\psi \,. \label{Dpsi=apsi}
\end{equation}
Multiplying both sides by $\Gamma^\mu$, this reduces to the Dirac equation where the mass $m=3\alpha$ is an integration constant related to the torsion of the background, in complete agreement with \cite{AVZ}.

Both (\ref{T=ee}) and (\ref{Dpsi=apsi}) break local scale invariance, leaving only a global symmetry under $e^a\rightarrow \lambda e^a$, $\psi \rightarrow \lambda^{-1}\psi$, $m \rightarrow\lambda^{-1}m$ for constant $\lambda$. In analogy with SUSY, this rigid symmetry does not interfere with the counting of local DOF. As pointed out in \cite{AVZ,SU2}, the introduction of a dimensionful mass constant $m$ enables us to finally determine the scale for the theory.

In Appendix \ref{solving_consistency}, we show that the sector equations \eqref{T=ee} and \eqref{Dpsi=apsi} can be used to consistently solve and preserve the remaining constraints. In fact, in this case one is enabled to explicitly determine the time evolution of $e$ and $\psi$, which is equivalent to the fact that Lagrange multipliers in the total Hamiltonian are also found in closed form (without using the degenerate gauge \eqref{ea=xiei}). We now show how the first-class generators arise to recover the residual symmetries of (\ref{T=ee},\ref{Dpsi=apsi}). In principle we will only assume the spatial components of these equations to hold, while the temporal parts will be recovered from Hamilton equations. Thus, note first that in this sector the combinations
\begin{eqnarray}
\tilde{K}_a & = & K_a -D_i \varphi^i_a + 2\alpha \epsilon^b{}_{ac} e^c_i \varphi^i_b +\alpha(\bar{\Omega}\Gamma_a \psi - \bar{\psi}\Gamma_a \Omega) \nonumber \\
& & +2ie^b_i \bar{\psi} \Gamma_{ab}\psi\phi^{i} + 2\epsilon^b{}_{ac} e^c_i \bar{\psi} \psi \phi^i_b \, , \label{tildeKa}
\end{eqnarray}
are first-class constraints, as can be directly checked computing the Poisson brackets:
\begin{eqnarray}
\{\tilde{K}_{a},\Omega\} \approx\{\tilde{K}_{a},\bar{\Omega}\} \approx\{\tilde{K}_{a},\varphi_{b}^{i}\} &  \approx & 0 \,, \\
\{\tilde{K}_{a},\phi_{b}^{i}\} \approx\{\tilde{K}_{a},\phi^{i}\} &  \approx & 0 \,, \\
\{\tilde{K}_{a}, \tilde{K}_{b} \} &\approx & 0 \,.
\end{eqnarray}
These three constraints are the generators of spacetime diffeomorphisms supplemented by gauge transformations and projected on the tangent space. This is seen from the identity
\begin{equation}
\{\cdots ,e^a_i \tilde{K}_a \} \approx \{ \cdots, \mathcal{H}_i + A_i \tilde{K} + \omega^a_i \tilde{J}_a \} \,.
\end{equation}
We now set the Lagrange multipliers associated to the primary constraints in order to accommodate the seven first-class generators. The total Hamiltonian reads
\begin{eqnarray}
H_T & = & \int d^{2}x \left[e^a_t K_a -\omega^a_t J_a - A_t K + \varphi^i_a \lambda^a_i + \phi^i_a \Lambda^a_i + \bar{\Lambda}\Omega +\bar{\Omega} \Lambda + \lambda_i \phi^i\right] \nonumber \\
 & = & \int d^{2}x \left[-\omega^a_t \tilde{J}_a - A_t \tilde{K} + e_t^a \tilde{K}_a \right]  =: \int d^2 x\:\mathcal{H} \, . \label{total_hamiltonian_sector}
\end{eqnarray}

Note that here we are implicitly fixing the Weyl freedom, i.e. we have assumed $v=0$ in the shift (\ref{Hshift}). This is required to preserve the sector. Indeed, the time evolution for the fields $(e^a_i,\psi,A_i,\omega^a_i)$, by virtue of the Hamilton equations, leads to
\begin{eqnarray}
D_t \psi=\dot{\psi} -\frac{i}{2}A_t \psi + \frac{1}{2}\epsilon^{abc}\omega_{bt} \Gamma_c \psi & = & \alpha\Gamma_t \psi\,, \label{D0psi}\\
T^a_{it}=\partial_i e^a_t -\dot{e}^a_i + \omega^a_{bi}e^b_t -\omega^a_{bt}e^a_i & = & 2\alpha\epsilon^a_{\:bc} e_i^b e_t^c\,, \label{Ti0}\\
F_{it} = \partial_i A_t -\dot{A}_i & = & 2ie_{i}^{a}e_t^{b}\bar{\psi}\Gamma_{ab}\psi\,, \label{Fi0}\\
R^a_{it} = \partial_i \omega^a_t -\dot{\omega}_i + \epsilon^{abc}\omega_{bi} \omega_{ct}  & = & 2\epsilon_{\:bc}^a e_i^b e_t ^c \bar{\psi}\psi\, , \label{Ri0}
\end{eqnarray}
These are readily seen to recover the temporal parts of equations (\ref{T=ee},\ref{Dpsi=apsi}) and the constrains (\ref{Ja},\ref{K}), thus agreeing with the Euler-Lagrange equations.

As stated, an interesting feature of this gauge is that $e^a_t$ is not restricted at all, which is equivalent to the statement that the three constraints $\tilde{K}_{a}$ are first-class. For regular configurations with $|e|\neq0$, it is clear that $(\mathcal{H},\mathcal{H}_i)$ are then three independent constraints generating temporal and spatial diffeomorphisms, respectively. Nevertheless, even for a degenerate vielbein it is possible to define
\begin{equation}
\mathcal{H}_{\perp}:=\epsilon_{\,bc}^{a}e_{1}^{b}e_{2}^{c}\tilde{K}_{a}\,,
\end{equation}
which corresponds (up to normalization) to the generator of diffeomorphisms normal to the surfaces  $t= constant$, modulo gauge transformations. Defining the Lagrange multipliers $e^a_t$, $A_t$ and $\omega^a_t$ as
\begin{eqnarray}
e^a_t & = & N^{\perp}\epsilon^a{}_{bc} e^b_1 e^c_2 + e^a_i N^i \,,\\
A_t & = & \lambda- A_{i}N^i \,,\\
\omega^a_t & = & \lambda^a - \omega^a{}_i N^i\,,
\end{eqnarray}
the generator of time evolution takes the more familiar form \cite{NT}
\begin{equation}
\mathcal{H}= N^{\perp}\mathcal{H}_{\perp} + N^{i}\mathcal{H}_i - \lambda\tilde{K}-\lambda^{a}\tilde{J}_{a}.
\end{equation}

We thus find the expected $SO(2,1)\times U(1)\times \text{Diff} $  residual symmetries and their corresponding generators. We anticipate that even though in this gauge choice there exist a different set of first-class contraints associated to diffeomorphisms, the number of DOF is the same and this is therefore a generic sector. This will be discussed in Section \ref{degree_counting}.

\subsection{Bosonic Vacuum}\label{bosonic_vacuum}      
The purely bosonic vacuum $\bar{\psi}=0=\psi$ corresponds to a very particular configuration. In principle, it should \textit{not} be regarded as a subsector of the previous case because it acquires additional degeneracies in the Dirac matrix\footnote{The Dirac matrix is defined as $\Omega_{AB}:=\{\phi_A,\phi_B\}$, where the indexes $A,B$ range over all the constraints \cite{Dirac}.} which lead to new first-class constraints. This is a direct consequence of the whole energy-momentum tensor of the Lagrangian formalism vanishes identically and there are no field equations to determine $e^a_\mu$, so the dreibein is a non-propagating gauge field in this case. Nevertheless, some of the first-class constraints found in the previous section turn out to be not functionally independent and therefore compensate the situation. As we will show, the whole picture results into an equal number of DOF, thus we can think of the vacuum as a generic sector.

First note if the fermions vanish, \eqref{prim-prim} and \eqref{Ka}-\eqref{K} imply
\begin{eqnarray}\label{vacphi}
\{\varphi_{a}^{i},\varphi_{b}^{j}\} & = & \{\varphi_{a}^{i},\Omega\}=\{\varphi_{a}^{i},\bar{\Omega}\}=\{\varphi_{a}^{i},\phi^{j}\}=\{\varphi_{a}^{i},\phi_{b}^{j}\}=0 \,,\\
\{\varphi_{a}^{i},K\} & = & \{\varphi_{a}^{i},J_{a}\}=\{\varphi_{a}^{i},K_{a}\}=0 \, ,
\end{eqnarray}
(where we have set $\bar{\psi}=0=\psi$ \textit{after} computing the Poisson brackets). Thus, we find six additional first-class constraints $\varphi^i_a \approx 0$, which generate arbitrary changes in the spatial components of the dreibein,
\begin{equation}\label{deltae}
\delta e_{i}^{a}=\{e_{i}^{a},\int d^{2}x\:\lambda_{j}^{b}\varphi_{b}^{j}\}=\lambda_{i}^{a}\,.
\end{equation}
As the time component $e^a_t$ is already a Lagrange multiplier, this in turn means that the dreibein is completely arbitrary (in particular it can be chosen to be invertible). In this sector, the first-class constraints \eqref{first_class_constraints} read
\begin{eqnarray}
\tilde{J}_{a} & = & J_{a}+\epsilon_{\:ac}^{b}\varphi_{b}^{j}e_{j}^{c}+D_{j}\phi_{a}^{j}\label{vacJa}  \,, \\
\tilde{K} & = & K+\partial_{j}\phi^{j} \label{vacK}  \,, \\
\Upsilon & = & -e_{j}^{b}\varphi_{b}^{j} \label{vacY}  \,, \\
\mathcal{H}_{i} & = & e_{i}^{a}K_{a}-e_{i}^{a}D_{j}\varphi_{a}^{j}+T_{ij}^{a}\varphi_{a}^{j} -\omega_{i}^{a}\tilde{J}_{a}-A_{i}\tilde{K} \,. \label{vacHi}
\end{eqnarray}
Note that the Weyl invariance has not been fixed so the torsion components $T^a_{ij}$ remain undetermined. In this sector one can also identify $K_a \approx 0$ as a first-class constraint (which is identically fulfilled). However, since \eqref{Ka} is quadratic in the fermionic variables, it can be shown that it does not act on the phase space,
\begin{equation}\label{Katrivial}
\{K_{a},F\}=0\,,
\end{equation}
for any function of the physical fields. As $K_a := \frac{\partial \mathcal{L}}{\partial e^a_t}$ can be regarded as the ($t-a$) components of the energy-momentum tensor, \eqref{Katrivial} is a consequence of the fact that the linearized version of $\mathcal{T}^\mu_{\,\,\nu}=0$ is fulfilled identically. Considering this functional degeneracy of $K_a$, we see that diffeomorphisms (\ref{vacHi}) are composed only of gauge transformations plus certain particular displacements of the vielbein. Moreover, it is clear that the $SO(2,1)\times U(1)\times \text{Diff}\times Weyl$ transformations are generated by a linear combination of the first-class constraints $\tilde{J}_a ,\tilde{K}$ and $\varphi_{a}^{i}$ only. The remaining constraints, corresponding to $\Omega$, $\bar{\Omega}$, $\phi^j_b$ and $\phi^j$, are second-class as can be checked from their Poisson brackets \eqref{prim-prim}.

\subsubsection{Dirac brackets and charge algebra}  

As the rank of the Dirac matrix is constant for a neighborhood of the vacuum in the phase space (the generic property is defined for open regions \cite{BGH}), the above classification of constraints can in fact be applied for a small perturbation with $\psi \neq 0$. We now illustrate this by computing the Dirac brackets. According to \eqref{prim-prim} and the definition of Dirac brackets \cite{Dirac}, one finds
\begin{eqnarray}\label{Dirac_bra}
\{A_i,A_j \}_D & = & \epsilon_{ij} ,\quad
\{\omega^a_i,\omega^b_j\}_D = \epsilon_{ij}\eta^{ab} , \quad
\{\bar{\psi}_\beta,\psi^\alpha \}_D  =  \hat{\Gamma}^\alpha_{\,\, \, \beta} ,\nonumber \\
\{e^a_i,p^j_b \}_D & = & \delta^j_i\delta^a_b , \quad
\{p^j_b,\psi \}_D  = 2\epsilon^{ij}e^a_i \hat{\Gamma}\Gamma_{ab} \psi ,  \quad  \{p^{j}_{b},\bar{\psi}\}_D = 2\epsilon^{ij}e^a_i \bar{\psi}\hat{\Gamma}\Gamma_{ab}   \, ,
\end{eqnarray}
where $\epsilon_{ij}\epsilon^{ik}=\delta^k_j$ and the matrix $\hat{\Gamma}$ is defined such that
\begin{equation}
2\epsilon^{ij}(\Gamma_{ij})^\alpha_{\,\, \, \beta}(\hat{\Gamma})^\beta_{\,\, \, \gamma}=\delta^\alpha_\gamma \, ,
\end{equation}
(explicitly, $\hat{\Gamma}=-\frac{1}{4|e|g^{tt}}E^t_a\Gamma^a$). Note that the phase space reduces to the fields $(A_i, \omega^a_i, \psi, \bar \psi, e^a_i, p^j_b)$ after the second-class constraints are strongly implemented. With this simplification the first-class generators \eqref{first_class_constraints} read
\begin{eqnarray}
\tilde{J}_{a} & = & J_{a}+\epsilon_{\:ac}^{b}\varphi_{b}^{j}e_{j}^{c} \label{DJa}  \,, \\
\tilde{K} & = & K  \label{DK}  \,, \\
\Upsilon & = & -e_{j}^{b}\varphi_{b}^{j} = -e_{j}^{b}p_{b}^{j}  \label{DY}  \,, \\
G_i:= \mathcal{H}_{i} + \omega_{i}^{a}\tilde{J}_{a} + A_{i}\tilde{K}  & = & e_{i}^{a}K_{a}-e_{i}^{a}D_{j}\varphi_{a}^{j}+T_{ij}^{a}\varphi_{a}^{j} \,. \label{DHi}
\end{eqnarray}
where $G_i$ is the generator of \textit{improved} diffeomorphisms \cite{BGH} (under the Weyl fixing of Section \ref{half_spin_sector} it simply reads $G_i=e^a_i \tilde{K}_a$). As we will show, under appropriate boundary conditions $G_i$ does not contribute to asymptotic charges, i.e. it corresponds to proper gauge transformations. Following the Regge-Teitelboim approach \cite{RT}, the smeared gauge generator must be supplemented by a boundary term $Q$ depending on the asymptotic gauge parameters, so it reads
\begin{eqnarray}\label{smear}
S[\xi^i, \lambda, \lambda^a,v]=\int d^2x (\xi^i G_i + \lambda \tilde{K} + \lambda^a \tilde{J}_a + v\Upsilon)  + Q_G[\xi^i]+Q_{\tilde{K}}[\lambda]+Q_{\tilde{J}}[\lambda^a] \, .
\end{eqnarray}
These boundary terms correspond to the asymptotic charges associated to (global) gauge symmetries\footnote{These conserved charges are determined
by the boundary terms that must be added to the action \eqref{I} in order to have well defined functional derivatives with respect to the fields. In \cite{Banados}, for instance, the normalization factor is chosen to be $\frac{k}{4\pi}$, where $k$ is the CS level.}. As stated above, the Weyl scaling does not have an associated charge. In Appendix \ref{apcharges} we give the form of the variation of the charges and integrate them. It is shown that setting the fermionic fields $(\psi, \bar \psi)$ to vanish asymptotically\footnote{This condition is preserved under all the gauge symmetries considered here (see Appendix \ref{apcharges} for details). However, regarding supersymmetry, one needs to check the stability under such transformation by solving the Killing spinor equation for a certain background, as shown in \cite{AVZ} for the BTZ case.} yields no boundary term $ Q_G[\xi^i]$, so the conserved charge due to diffeomorphisms $\mathcal{H}_i$ is solely due to $SO(2,1) \times U(1)$ gauge transformations, as usual for CS systems in $2+1$ dimensions \cite{Banados}.

Once the boundary terms have been determined one is able to recover the gauge transformations globally generated by $S$ under the Dirac bracket. For instance, direct computation yields explicit relations for the improved diffeomorphisms (\ref{D1},\ref{Domega}). As can be readily checked, the smeared constraints $\tilde{K}$, $\tilde{J}_a$ and $\Upsilon$ also yield the corresponding transformations. Finally, as shown in Appendix \ref{apcharges}, the asymptotic algebra induced by these symmetries splits into a (local) direct product $SO(2,1) \times U(1) $ with the corresponding central extensions:
\begin{eqnarray}\label{asymptoticalg}
\{ Q_{\tilde{K}}[\lambda],  Q_{\tilde{K}}[\zeta] \}_D &=& C_{\tilde{K}}[\lambda,\zeta] \, , \\
\{ Q_{\tilde{J}}[\lambda^a],  Q_{\tilde{J}}[\zeta^a] \}_D &=& Q_{\tilde{J}}[\epsilon^a_{\,\, bc}\lambda^b \zeta^c]+C_{\tilde{J}}[\lambda^a,\zeta^a] \, ,
\end{eqnarray}
where the central terms $C_{\tilde{K}}$ and $C_{\tilde{J}}$ given in (\ref{asymptU1},\ref{asymptSOU1}), do not depend on the dynamical fields but only on the gauge parameters. A further refinement of this algebra can be obtained by the well known procedure of imposing asymptotic conditions for the bosonic sector $(\omega_\mu, A_\mu)$ \cite{Blagojevic}.

\section{Degree of freedom count}\label{degree_freedom_section}      
\label{degree_counting}
In a theory with $N$ dynamical field components (that is, excluding Lagrange multipliers), $F$ first-class and $S$ second-class constraints, the number of DOF is given by \cite{HTZ}
\begin{equation} \label{g}
g= \dfrac{2N-2F-S}{2}\, .
\end{equation}

In the system discussed here there are $N=14+4n$ dynamical field components, $A_i, \omega^a{}_{i}, e^a_i, \psi, \bar{\psi}$.  The following table gives  the values of $F$ and $S$ in different cases:
\begin{center}
\begin{tabular}{|c|c|c|c|c|}
\hline
Sector & Gauge & Generators & $\quad F \quad$ & $S$  \tabularnewline
\hline
\hline
Any generic & $N^{\perp} = 0$ &$\tilde{J}_{a},\tilde{K},\mathcal{H}_i,\Upsilon$ & 7 & $14+4n$ \tabularnewline
\hline
Spin-1/2 & $v=0$ &$\tilde{J}_{a},\tilde{K},\tilde{K}_{a}$ & 7 & $14+4n$ \tabularnewline
\hline
Vacuum & --- &$\tilde{J}_{a},\tilde{K},\varphi_{a}^{i}$ & 10 & $8+4n$ \tabularnewline
\hline
\end{tabular}
\end{center}
In all cases, formula (\ref{g}) gives $g=2n$, in complete agreement with the naive counting of Section \ref{introduction}. Note that the first two sectors share the same number of independent first-class constraints. For the second, one finds an additional diffeomorphism generator instead of the Weyl scaling.

As the possibility of finding another first-class combination cannot be ruled out in general, one could in principle find a sector where all the three diffeomorphism generators and the Weyl scaling (in addition to $\tilde J_a$ and $\tilde K$) are independent, even though such a configuration would certainly be \textit{non-generic} by definition. However, this would lead to an odd number of second-class constraints and a non-integer result for $g$, according to \eqref{g}.
\section{Discussion and summary}  

We have carried out the Dirac analysis for constrained Hamiltonian systems for the action composed of a spin-$1/2$ Dirac field minimally coupled to an electromagnetic potential and to the Lorentz connection in $(2+1)$-dimensions. The action of the entire system (\ref{I}) is obtained from a CS form for an $\mathfrak{osp}(2|2)$ connection, in which the spinorial component of the connection was split as $\chi^\alpha_\mu := e^a_\mu(\Gamma_a)^\alpha_\beta \psi^\beta$. This splitting has a number of nontrivial consequences for the dynamical contents of the theory: i) Instead of zero degrees of freedom of a generic CS action, this system has the four propagating DOF of a Dirac spinor; ii) The system acquires a \textit{proper} Weyl rescaling symmetry, i.e., it has no associated Noether charge and can be directly fixed; iii) The metric structures --the dreibein and the induced metric-- are invariant under SUSY, and therefore there is no need to include spin-3/2 fields (gravitini); iv) Supersymmetry is reduced from a gauge symmetry to a rigid/global invariance that is contingent on the features of the background geometry and the gauge fields; v) For the vacuum sector the dreibein becomes pure gauge and diffeomorphisms degenerate into $SO(2,1) \times U(1)$ transformations.

The Dirac formalism completely recovers the Lagrangian equations. The equations for the gauge fields $(\omega, A)$ follow from the constraints and the Hamilton equations for these fields. Furthermore, it can be shown that the Dirac equation and equation $\mathcal{T}^\mu{}_\nu=0$ are respectively equivalent to (\ref{reduced1},\ref{reduced2}) for an invertible dreibein. In fact, after Weyl fixing and computing the temporal evolution one gets $D_t \psi = e^a_t \zeta_a$ and $T^a_{ti}= e^b_t e^c_i T^a_{\,\,\, bc}$. Then, equations (\ref{reduced1},\ref{reduced2}) together with the constraint \eqref{Ka} can be covariantized to give
\begin{eqnarray}\label{covariantC4C5}
T^a_{\mu \nu} \bar{\psi} \psi  &=& \bar{\psi}\Gamma^a \Gamma_{[\mu} \overrightarrow{D}_{\nu]} \psi + \bar{\psi} \overleftarrow{D}_{[\nu} \Gamma_{\mu]} \Gamma^a \psi \, , \label{covc4}\\
\Gamma^{\mu} D_{\mu}\psi &=& \frac{1}{4} T^a_{\mu \nu}\Gamma^{\nu \mu} \Gamma_a \psi \,.
\end{eqnarray}
The degeneracy of these equations follows from the fact that $\mathcal{T}^\mu_{\,\,\,\mu}$ is proportional to \eqref{covc4} and is a combination of the Dirac equation -plus its conjugate-, and therefore identically vanishes for this theory, which is in turn equivalent to Weyl invariance.

It should be stressed that $g=2n$ is an upper bound for the number of local DOF, since in non-generic sectors there might be additional accidental first-class constraints and therefore fewer degrees of freedom, as it happens in some sectors of higher-dimensional CS systems \cite{BGH}. The general counting performed in Section \ref{generic_sector} proceeds under the assumption that this is not the case. The argument given there, using the degenerate gauge, even holds for the spin-$1/2$ sector of Section \ref{half_spin_sector}, but for that configuration it is illustrative to explicitly use the Weyl fixing instead (see the end of Appendix \ref{solving_consistency}).

In that sense, the purpose of choosing a specific sector such as the spin-$1/2$ is twofold: On the one hand the Lagrange multipliers can be readily solved, allowing for an explicit solution of (\ref{reduced1},\ref{reduced2}) leading to a full realization of the first-class constraints. On the other, the Weyl symmetry is ``gauged away" in this case, providing a symmetry breaking mechanism. One is left with a global version of the scale invariance which is broken by fixing the fermion mass or the normalization of the dreibein.

In this system SUSY seems to play a marginal role. It starts out as part of the gauge invariance of the action (\ref{I}), then it is seen as a global (rigid) symmetry without first-class constraints associated to it, contingent on the existence of some spacetime symmetry, which need not occur in every spacetime background.
The action and the equations are invariant under
\begin{eqnarray}\label{susy}
\delta \psi & = & \frac{1}{3}\slashed{D} \epsilon \, , \quad \delta \overline{\psi} = \frac{1}{3}\overline{\epsilon}\overleftarrow{\slashed{D} }\nonumber \\
\delta A & = & -\frac{i}{2}\left( \overline{\psi}\slashed{e} \epsilon +\overline{\epsilon}\slashed{e} \psi \right)			\nonumber \\
\delta \omega^a & = &-\overline{\psi}\left( e^a+ \epsilon^a{}_{bc}e^b \Gamma^c \right) \epsilon - \overline{\epsilon} \left( e^a - \epsilon^a{}_{bc}e^b \Gamma^c \right)\psi \, , \\
\delta e^a & = & 0\, \nonumber
\end{eqnarray}
where $\epsilon$ satisfies the no-spin-3/2 condition, $[\delta_\nu ^\mu - (1/3) \Gamma_\nu \Gamma^\mu ]D_\mu \epsilon =0$. This condition can be fulfilled provided the spacetime and the connection fields admit a Killing spinor of a certain kind \cite{APZ}. This is the case for the vacuum: AdS or Minkowski space without fermions or electromagnetic fields. This background is a full-BPS state preserving full supersymmetry, but there are configurations preserving 1/2 or 1/4 of SUSY, just like in 2+1 supergravity \cite{Coussaert-Henneaux,SU2}. A bosonic vacuum $\psi=0$ remains invariant under (\ref{susy}) provided $\slashed{D} \epsilon =0$, which is also a requirement that the background admit a Killing spinor.

This unconventional SUSY can by extended to describe fermions in the fundamental representation of a non-Abelian internal group like $SU(2)$ \cite{SU2}. As we have seen above, the constraints associated to internal $U(1)$ and Lorentz $SO(2,1)$ symmetries decouple from the diffeomorphism and Weyl ones. By the same token, in a generic supersymmetric extension of an internal non-Abelian gauge symmetry, the fermion excitations turn out to be the only contribution to the local DOF. In order to illustrate this, let us consider the split Lagrangian for the $SU(2)$ theory, which, up to a global factor, reads \cite{SU2}\footnote{The CS form also contains an abelian form $b$ associated to the central charge in $su(2,1|2)$. However, $b$ decouples from the action and therefore does not enter in the dynamical analysis.}
\begin{eqnarray}
L_{SU(2) }& = & \epsilon^{ij}\Bigl[-\eta_{ab} \dot{e}_i ^a e_j^b \bar{\psi}_A \psi^A -\dot{\bar{\psi}}_A\Gamma_{ij}\psi^A+\bar{\psi}_A\Gamma_{ij}\dot{\psi}^A +\frac{1}{2}\eta_{ab}\dot{\omega}_i^a \omega_j ^b  + \frac{1}{2}\delta_{IJ}\dot{A}^I_i A^J_j \Bigr] \nonumber \\
& & -e_t^a K_a + \omega_t ^a J_a +A^I_t K_I\,,\label{eq:SplitlagSU2}
\end{eqnarray}

Here the indexes $A=1,2$ transform under the $2\times2$ vector representation of $SU(2)$ (Pauli matrices), while $I=1,2,3$ refers to the adjoint representation (we follow the conventions of \citep{SU2}). The primary constraints $(\varphi^i_a, \phi^i_a, \phi^i_I, \Omega^A, \bar{\Omega}_A)$ are defined in an analogous fashion to their $U(1)$ counterparts. If one omits the contraction in the $A$ index, i.e. $\bar{\psi}_A \psi^A = \bar{\psi} \psi$, the secondary constraints $K_a$ and $J_a$ adopt exactly the same form as (\ref{Ka},\ref{Ja}) where the covariant derivatives are now gauged by $SO(2,1) \times SU(2)$. The remaining constraint reads
\begin{equation}\label{KISU2}
K_{I} =\epsilon^{ij}\delta_{IJ}(\frac{1}{2}F^{J}_{ij}-i\bar{\psi}\Gamma_{ij}\sigma^{J}\psi) = \epsilon^{ij}\delta_{IJ}(\partial_{i}A_{j}^{J}+\frac{1}{2}\epsilon_{\,KL}^{J}A_{i}^{K}A_{j}^{L}-i\bar{\psi}\Gamma_{ij}\sigma^{J}\psi) \, .
\end{equation}
Then, one can show that
\begin{eqnarray}\label{first_class_SU2}
\tilde{J}_a & := & J_a +\epsilon^{}_{ac} \varphi^j_b e^c_j +\frac{1}{2}(\bar{\Omega}\Gamma_a \psi - \bar{\psi}\Gamma_a \Omega) + D_j \phi^j_a \,, \nonumber \\
\tilde{K}_I & := & K_I-\frac{i}{2}(\bar{\Omega}\sigma_I \psi-\bar{\psi}\sigma_I\Omega)+D_j \phi^j_I \,,\nonumber \\
\Upsilon & := & -e^b_j \varphi^j_b +\bar{\Omega}\psi+\bar{\psi}\Omega \,, \\
\mathcal{H}_i & := & e^a_i K_a -e^a_i D_j \varphi^j_a + T^a_{ij} \varphi^j_a + \bar{\psi} \overleftarrow{D}_i \Omega +\bar{\Omega}\overrightarrow{D}_{i}\psi-\omega_{i}^{a}\tilde{J}_{a}-A^I_{i}\tilde{K}_I+\phi^{j}_I F^I_{ij} +\phi_{a}^{j}R^a_{ij} \, . \nonumber
\end{eqnarray}
correspond to $F=9$ first-class combinations generating $SO(2,1) \times SU(2) \times Weyl \times \text{Diff}$ transformations, respectively. In account of these and the remaining $S=18+8n$ second class constraints, the original phase space of $N=18+8n$ variables only contains
\begin{equation}\label{gSU2}
g_{SU(2)}= \dfrac{2N-2F-S}{2} = 4n \,
\end{equation}
degrees of freedom for a generic sector, exactly matching the double of the $U(1)$ case due to the doubling of the fermion fields. SUSY is again not realized as a first-class constraint, but is a rigid transformation for certain backgrounds. Such matters, together with the computation of the asymptotic charges, were already treated in the original work.

Unconventional supersymmetries can also be constructed in higher dimensions based on a gauge superalgebra containing $\mathfrak{so}(2n,2)$ or $\mathfrak{so}(2n-1,1)$ as a proper subalgebras. In odd dimensions $D=2n+1\geq 5$, a similar CS construction can be set up, while for $D=2n \geq 4$, since the CS forms are not defined, the construction requires a metric and the action can be of a Yang-Mills type. In both cases the fermionic part of the connection can be construed as a composite of a vielbein and a spin-1/2 Dirac field \cite{APZ}. For all $D\geq4$, it can be expected that, as in the three-dimensional case discussed here, the vielbein would not contribute to the dynamic contents unless it possesses an independent kinetic term of its own; the effective gauge symmetry would correspond to the bosonic part of the superalgebra, and supersymmetry would be reduced to a rigid invariance conditioned by the existence of globally defined Killing spinors of the background. In other words, supersymmetry would be at most an approximate feature in some vacuum spacetime geometries, and the main footprint of its presence in the theory would be in the field content, the type of couplings and the parameters in the action.

The conduction properties of graphene \cite{Wallace,Semenoff,Novoselov} can be very well described by the $\pi$-electrons in the two sublattices of the honeycomb structure as massless fermions in the long-wavelength limit \cite{graph_rev}. It was already conjectured that the system studied here could reproduce the behavior of these $\pi$-electrons \cite{AVZ,SU2}, while the very strong $\sigma$-bond of the remaining available electrons of the carbon atoms keep the geometry of the graphene layer fixed. Therefore it is expected that in the low energy (long wavelength) regime, the dynamical contents are essentially in the fermion sector, as pointed out here. Nevertheless, note that we have introduced a torsional mass term, which is required in principle by hermiticity. Such construction not only leads to a symmetry breaking mechanism but, it also allows the massive fermion to trigger a backreaction into the background, provided we use the contorsion as an effective cosmological constant. This implies a constant curvature background as illustrated in \cite{AVZ}. Following that line, an idea to be experimentally explored is whether specific graphene layers (or graphene-like material) can be manufactured which admit Killing-spinors in order to measure some induced supersymmetric effects. This would provide low-energy graphene models to test high-energy physics theories, whose observable effects are beyond reach in current particle accelerators \cite{Iorio}.

Besides providing a rigorous tool for identifying the dynamical DOF, the Hamiltonian formalism could be the preliminary warm-up towards a quantization procedure \cite{HT,Dirac}, eventually leading to a quantum theory of graphene. In the system described here, the only dynamical degrees of freedom are those of the Dirac fermion; the bosonic connections $A$ and $\omega$ are described by Chern-Simons actions and therefore have no local degrees of freedom, while the dreibein is an artifact that can be gauged away. This means that the bosonic fields do not contribute to the quantum field theory other than as classical external fields; their quantum excitations are produced by nontrivial global holonomies of a topological nature. Such fields do not propagate and hence do not generate perturbative corrections. In particular, there should be no perturbative corrections generated by quantum fluctuations of the bosonic fields in graphene, the system should behave like a free electron field propagating in a curved classical background and would therefore be renormalizable.

Further insight comes from the AdS$_3$/CFT$_2$ duality and its generalizations in $2+1$ gravity \cite{Brown, Banados,CHvD}, which are realized through a centrally extended \textit{canonical} algebra in the asymptotic region associated to a quantum theory at the boundary. In the broader gauge/gravity context, the holographic description of graphene in the IR regime has been recently studied by means of a 3+1 D-brane embedding, exhibiting also conformal symmetry breaking due to the introduction of a mass gap scale as an integration constant \cite{Evans}.

\section*{Acknowledgments}
We are grateful to P.~D.~Alvarez, W.~Clemens, O.~Fuentealba, J.~Helayel-Neto, A.~Iorio, F.~Toppan and M.~Valenzuela for many enlightening discussions. We wish to specially thank P. Salgado-Rebolledo for taking active part in the initial stages of this project. P. P. wishes to express his thanks to Prof. M.~Henneaux for support and encouragement. A. G. also thanks CONICYT for financial support. This work has been partially funded through Fondecyt grant 1140155. The Centro de Estudios Cient\'{\i}ficos (CECs) is funded by the Chilean Government through the Centers of Excellence Base Financing Program of Conicyt.

\appendix
\section{General definitions and useful properties}\label{conventions}   
Through this work we extensively use the Clifford algebra in $D=3$. Some basic properties and definitions are\footnote{We adopt the convention $\epsilon_{012}= -\epsilon^{012}=1$ and the definition $T_{[a_1...a_p]}=\frac{1}{p!}\delta_{a_1...a_p}^{b_1...b_p}T_{b_1...b_p}$. In the coordinate basis, $\epsilon_{ij}:=\epsilon_{tij}$.}
\begin{eqnarray*}
\{\Gamma_a,\Gamma_b\} = 2\eta_{ab}\,, \quad  \Gamma_{ab}:=\Gamma_{[a}\Gamma_{b]}  =  \frac{1}{2}[\Gamma_{a},\Gamma_{b}]\, , \quad \Gamma_{ab} =  \epsilon_{abc}\Gamma^{c} \,, \quad \\
\quad \frac{1}{2} [  \Gamma_{ab},\Gamma_c ] =\eta_{bc}\Gamma_a -\eta_{ac} \Gamma_b \,, \quad  \Gamma_{abc}  =  \epsilon_{abc} =
\frac{1}{2}\{\Gamma_{ab},\Gamma_{c}\}=\Gamma_{[a|}\Gamma_b \Gamma_{|c]}
\end{eqnarray*}
Let $\psi$ be a two-component Dirac spinor with Grassman parity odd. We define its Dirac conjugate by
\begin{equation}\label{dirac_conjugate}
\overline{\psi}=i\psi^{\dagger}\Gamma_0.
\end{equation}
or explicitly as $\bar \psi_\beta = i \psi^{\alpha * }(\Gamma_0)_{ \alpha \beta}$, $\alpha, \beta = 1,2$.
With this prescription, we have the conjugacy properties
\begin{eqnarray}\label{reality_conditions}
(\overline{\chi}\psi)^{\ast}&=&\overline{\psi}\chi, \nonumber \\
(\overline{\chi}\Gamma_{a}\psi)^{\ast}&=&-(\overline{\psi}\Gamma_{a}\chi). \\
\overline{\left(\Gamma_{a}\psi\right)} & = & -\overline {\psi}\Gamma_{a} \nonumber
\end{eqnarray}

The starting point for the this model is to take the connection for the full $osp(2|2)$ algebra \cite{AVZ}
\begin{equation}\label{connection}
\mathcal{A}=A\mathds{Z}+\omega^{a}\mathds{J}_{a}+\mathds{\overline{Q}}\slashed{e}\psi+\overline{\psi}\slashed{e}\mathds{Q},
\end{equation}
where $\mathds{Z}$ is the $U(1)$ generator, $\{\mathds{J}_{a}\}$ is the set of generators of the Lorentz algebra $SO(2,1)$, and $\{\mathds{\overline{Q}}_{\alpha},\mathds{Q}^{\alpha}\}$ is the set of SUSY generators. The symbol $\slashed{e}$ is a compressed notation for $\slashed{e}=e^{a}\Gamma_{a}$. The algebra reads
\begin{eqnarray}\label{algebra}
\left[\mathds{J}_a,\mathds{J}_{b}\right] &=& \epsilon_{abc}\mathds{J}^{c}, \quad \left[\mathds{J}_{a},\mathds{Q}^{\alpha}\right] = -\frac{1}{2}(\Gamma_a)^{\alpha}_{\beta}\mathds{Q}^{\beta}, \quad \left[\mathds{J}_{a},\mathds{\overline{Q}}_{\alpha}\right] = \frac{1}{2}(\Gamma_{a})^{\beta}_{\alpha}\mathds{\overline{Q}}_{\beta}, \nonumber \\
\left[\mathds{Z},\mathds{Q}^{\alpha}\right] &=& \frac{i}{2}\mathds{Q}^{\alpha}, \quad \left[\mathds{Z},\mathds{\overline{Q}}_\alpha \right] = -\frac{i}{2}\mathds{\overline{Q}}_{\alpha}, \quad \left\{\mathds{Q}^{\alpha}, \mathds{\overline{Q}}_\beta \right\} =(\Gamma^{a})^{\alpha}_{\beta}\mathds{J}_{a}-i\delta^{\alpha}_{\beta}\mathds{Z},
\end{eqnarray}
where the other (anti-)commutators are zero.

The action \eqref{I} is given by
\begin{equation}
I[\psi,e,A,\omega]=\int \langle \mathcal{A},d\mathcal{A} \rangle  + \frac{2}{3} \langle \mathcal{A}, \mathcal{A}^2 \rangle
\end{equation}
where the (super-)invariants traces are
\begin{equation}\label{traces}
\langle\mathds{J}_a , \mathds{J}_b \rangle =\frac{1}{2}\eta_{ab}, \quad \langle\mathds{Z}, \mathds{Z}\rangle =\frac{1}{2}, \quad
\langle\mathds{\overline{Q}}_\alpha , \mathds{Q}^\beta \rangle = \delta^{\beta}_{\alpha},  \quad \langle\mathds{Q}^\alpha , \mathds{\overline{Q}}_\beta \rangle = -\delta^{\alpha}_{\beta}.
\end{equation}

The covariant derivative $D_{\mu}$ induced by \eqref{algebra} appears naturally in \eqref{I}. Acting on a Lorentz vector $\Sigma_{a}$ and $1/2$-spinors ($\psi^{\alpha}$ and $\overline{\psi}_{\alpha}$) this reads
\begin{eqnarray}\label{def_covariant_derivative}
D_{\mu}\Sigma_{a}&=&\partial_{\mu}\Sigma_{a}+\tensor{\epsilon}{_{ab}^{c}}\omega^{b}_{\mu}\Sigma_{c}, \nonumber \\
\overrightarrow{D}_{\mu}\psi^{\alpha}&=&\partial_{\mu}\psi^{\alpha}-\frac{i}{2}A_{\mu}\psi^{\alpha}+\frac{1}{2}\omega^{a}_{\mu}(\Gamma_{a})^{\alpha}_{\beta}\psi^{\beta}, \\
\overline{\psi}_{\alpha}\overleftarrow{D}_{\mu}&=&\partial_{\mu}\overline{\psi}_{\alpha}+\frac{i}{2}A_{\mu}\overline{\psi}_{\alpha}-\frac{1}{2}\overline{\psi}_{\beta}(\Gamma_{a})^{\beta}_{\alpha}\omega^{a}_{\mu}= \overline {(\overrightarrow{D}_{\mu}\psi )}_{\alpha}. \nonumber
\end{eqnarray}

\section{Momenta, Constraints and Poisson brackets}\label{poisson_brackets}   
For the starting action \eqref{eq:Splitlag}, the canonical momenta associated to the dynamical fields are given by
\begin{eqnarray}\label{canonical_momenta2}
\pi^{i}\approx \frac{\partial\mathcal{L}}{\partial\dot{A}_{i}}=\frac{1}{2}\epsilon^{ij}A_{j}, \quad  \pi^{i}_{a}\approx \frac{\partial\mathcal{L}}{\partial\dot{\omega}^{a}_{i}} &=& \frac{1}{2}\epsilon^{ij}\eta_{ab}\omega^{b}_{j}, \quad  p^{i}_{a}\approx \frac{\partial\mathcal{L}}{\partial\dot{e}^{a}_{i}} =-2\epsilon^{ij}\eta_{ab}e^{b}_{j}\overline{\psi}\psi \nonumber\\
\chi^{\alpha}\approx \frac{\partial^{L}\mathcal{L}}{\partial\dot{\overline{\psi}}_{\alpha}}=-\epsilon^{ij}(\Gamma_{ij})^{\alpha}_{\,\,\,\, \beta}\psi^{\beta} &,& \overline{\chi}_{\alpha}\approx \frac{\partial^{R}\mathcal{L}}{\partial\dot{\psi}^{\alpha}}=\epsilon^{ij}\overline{\psi}_{\beta}(\Gamma_{ij})^{\beta}_{\,\,\,\, \alpha}.
\end{eqnarray}
The non-vanishing Poisson brackets between the fields and their respective momenta are defined as in \cite{HT} \footnote{Hereafter we will omit the $\delta^2 (x-y)$ factors when computing the brackets. Spinor indexes may also be omitted for simplicity.}
\begin{eqnarray}\label{poisson-brackets}
\left\{A_i,\pi^j \right\} &=& -\left\{\pi^j, A_i \right\} = \delta^{j}_{i}, \quad  \left\{\omega^a_i, \pi^j_b \right\} = -\left\{\pi^j_b, \omega^a_i \right\} =  \left\{e^{a}_i,p^j_{b}\right\}=-\left\{p^j_{b},e^{a}_i\right\} =\delta^{j}_{i}\delta^{a}_{b}, \nonumber \\
\left\{\psi^\alpha, \overline{\chi}_\beta \right\} &=& \left\{\overline{\chi}_\beta, \psi^\alpha \right\} = \delta^\alpha_\beta, \quad
\left\{\overline{\psi}_{\alpha},\chi^{\beta}\right\}=\left\{\chi^{\beta},\overline{\psi}_{\alpha}\right\} =-\delta^{\beta}_{\alpha}.
\end{eqnarray}
It is worth to note the relative sign between the two brackets on the last line: This choice is consistent with $\overline{\psi}_{\alpha} = i\psi ^ {\beta *}(\Gamma_0)_{\beta \alpha } $ and $\overline{\chi}_{\alpha} = i\chi ^ {\beta *}(\Gamma_0)_{\beta \alpha } $. Now, the primary constraints \eqref{primary_constraints} satisfy
\begin{eqnarray}\label{prim-prim}
\{\varphi_{a}^{i},\varphi_{b}^{j}\}  = 4\epsilon^{ij}\eta_{ab}\bar{\psi}\psi ,\quad
\{\Omega,\varphi_b^j\} &=&2\epsilon^{ij}e_i^a \Gamma_a \Gamma_b \psi , \quad
\{\bar{\Omega},\varphi_{b}^{j}\}  =  2\epsilon^{ij}e_{i}^{a}\bar{\psi}\Gamma_{b}\Gamma_{a} ,\nonumber \\
\{\bar{\Omega},\Omega\}  = 2\epsilon^{ij}\Gamma_{ij} , \quad
\{\phi^{i},\phi^{j}\}  &=& -\epsilon^{ij} ,  \quad \{\phi_{a}^{i},\phi_{b}^{j}\} = -\epsilon^{ij}\eta_{ab}\,.
\end{eqnarray}

Using \eqref{prim-prim} and the definitions \eqref{first_class_constraints} one can show that the generators $\tilde{J}_{a}$, $\tilde{K}$ and $\Upsilon$ satisfy the following:
\begin{eqnarray}\label{sec-prim}
\{\tilde{J}_{a}, \phi_{b}^{i} \} &=&\epsilon_{ab}^{\quad c}\phi_c^i , \quad \{\tilde{J}_{a},\Omega\}=-\dfrac{1}{2}\Gamma_{a} \Omega , \quad \{\tilde{J}_{a},\bar{\Omega}\}=\frac{1}{2}\bar{\Omega}\Gamma_{a} ,  \quad \{\tilde{J}_{a}, \varphi_{b}^{i}\} = \epsilon_{ab}^{\quad c}\varphi_{c}^{i}\,, \nonumber \\
\{\tilde{J}_{a},J_{b}\} &=&\epsilon_{ab}^{\quad c}J_{c}\,,  \quad \{\tilde{J}_{a},K_{b}\}=\epsilon_{ab}^{\quad c}K_{c}, \quad  \{\tilde{K},\Omega\} =\frac{i}{2}\Omega ,  \quad  \{\tilde{K},\bar{\Omega}\}=-\frac{i}{2}\bar{\Omega} , \\
\{\Upsilon,\varphi_{a}^{j}\}&=& -\varphi_{a}^{j} ,  \quad  \{\Upsilon,K_{a}\}=-K_{a} , \quad \{\Upsilon,\Omega\}=\Omega , \quad \{\Upsilon,\bar{\Omega}\} = \bar{\Omega} , \nonumber
\end{eqnarray}
where the remaining brackets with constraints \eqref{Ka}-\eqref{primary_constraints} vanish strongly. We then conclude that the constraints  $\tilde{J}_{a}$, $\tilde{K}$,$\Upsilon$, together with the generator $\mathcal{H}_i$, are first-class.

The consistency of the primary constraints \eqref{primary_constraints} with respect to the extended Hamiltonian \eqref{hamiltonian} yields the following set of equations
\begin{eqnarray}\label{consistency}
0 = \left\{\phi^{i},H_{T}\right\}&=&\epsilon^{ij}\left(\partial_{j}A_t +2ie^{a}_t e^{b}_{i}\overline{\psi}\Gamma_{ab}\psi-\lambda_{j}\right)\,, \nonumber \\
0 = \left\{\phi^i_a ,H_T \right\}&=&\epsilon^{ij}\left(\eta_{ab}D_j \omega^b_t + 2\epsilon_{abc}e^b_t e^c_j \overline{\psi}\psi - \eta_{ab}\Lambda^b_j \right) \,, \nonumber \\
0 = \left\{\varphi^i_a, H_T \right\}&=&2\epsilon^{ij}\left(\epsilon_{abc}\omega^b_t e^c_j \overline{\psi}\psi -i A_t e^b_j \overline{\psi}\Gamma_{ab}\psi  -2 \eta_{ab}\partial_j \left(e^b_t \overline{\psi}\psi\right) -\epsilon_{abc}\omega^b_j e^c_t \overline{\psi}\psi \right. \\
&&\left. + e^b_t (\overline{\psi}\overleftarrow{D}_j \Gamma_a \Gamma_b \psi+\overline{\psi}\Gamma_b \Gamma_a\overrightarrow{D}_j \psi ) +2\eta_{ab}\lambda^b_j \overline{\psi} \psi + e^b_j \left(\overline{\Lambda}\Gamma_b \Gamma_a \psi + \overline{\psi}\Gamma_a \Gamma_b \Lambda\right) \right) \nonumber \,, \\
0 = \left\{\Omega ,H_T \right\}&=&-\epsilon^{ij}\left(iA_t e^a_i e^b_j \Gamma_{ab}\psi + \epsilon_{abc}\omega^a_t e^b_i e^c_j \psi + 2\eta_{ab}e^a_t T^{b}_{ij}\psi -2e^{a}_{t}e^{b}_{i}\Gamma_{a}\Gamma_{b}D_{j}\psi \right. \nonumber \\
&& \left. + 2 D_{j}\left(e^a_t e^b_i \Gamma_b \Gamma_{a}\psi \right) +2\lambda^{a}_{i}e^{b}_{j}\Gamma_{b}\Gamma_{a}\psi  -2e^{a}_{i}e^{b}_{j}\Gamma_{ab}\Lambda \right) \,, \nonumber \\
0 = \left\{\overline{\Omega} ,H_{T}\right\}&=&\epsilon^{ij}\left(iA_t e^{a}_{i}e^{b}_{j}\overline{\psi} \Gamma_{ab}- \epsilon_{abc}\omega^{a}_t e^{b}_{i}e^{c}_{j}\overline{\psi} - 2\eta_{ab}e^a_t T^{b}_{ij}\overline{\psi} +2e^{a}_t e^{b}_{i}(\overline{\psi}\overleftarrow{D}_{j})\Gamma_{b}\Gamma_{a}  \right. \nonumber \\
&&\left. - 2 \left(e^a_t e^b_i\overline{\psi}\Gamma_a \Gamma_b \right) \overleftarrow{D}_j -2\lambda^a_i e^b_j \overline{\psi}\Gamma_a \Gamma_b -2e^a_i e^b_j \overline{\Lambda}\Gamma_{ab} \right) \,. \nonumber
\end{eqnarray}
This system of $(14+4n)$ equations determines up to an equal number of Lagrange multipliers, leaving seven free parameters. This means that in a generic sector (maximum rank), there are $S=14+4n$ second-class and $F=7$ first-class constraints. Also, if one choose $(e^a_t,\omega^a_t,A_t)$ as the free parameters, the consistency of the secondary constraints $K_a, J_a$ and $K$ can be readily shown to follow. In Appendix \ref{solving_consistency} we exhibit a solution for \eqref{consistency}.

\section{Solving the consistency equations}\label{solving_consistency}     
Let us now choose tensors $\zeta_a$ and $T^a_{\,\,\, bc}=T^a_{\,\,\, [bc]}$, depending on the dynamical fields, such that
\begin{eqnarray}
T^a_{\,\,\, bc} e^b_i e^c_j &=&T^a_{ij}  \label{ansatzT} \\
&=& D_i  e^b_j - D_j e^b_i \,, \nonumber \\
e^a_i\zeta_a &=& D_i\psi \,. \label{ansatzD}
\end{eqnarray}

Equation (\ref{ansatzT}) relates the 9 Lorentz covariant components $T^a_{\,\,\, bc}$ to the 3 field dependent quantities on the RHS. Similarly, equation (\ref{ansatzD}) expresses the vector-spinor $\zeta_a$ as functions of 2 components on the RHS. This means there are six real and one spinorial indeterminate components respectively \footnote{For $|e|\neq 0$, one can put for instance $\zeta_a= E^i_a D_i\psi + E^t_a\xi$ and  $T^a_{\,\,\, bc}=E^i_b E^j_c T^a_{ij}+ E^t_{[b} E^i_{c]}\xi^a_i$ for arbitrary $\xi$ and $\xi^a_i$.}, which will be fixed by the consistency equations. Now, let us take the Lagrange multipliers in equation \eqref{hamiltonian} as

\begin{eqnarray}\label{fixmultipliers}
\lambda_{j}^{b} & = & -ve_{j}^{b}-\epsilon_{\:cd}^{b}\omega_t ^{c}e_{j}^{d}+D_{j}e_t ^{b}+T_{\,ac}^{b}e_t ^{a}e_{j}^{c}\,, \nonumber \\
\Lambda & = & v\psi+\frac{i}{2}A_t \psi-\frac{1}{2}\omega_t ^{c}\Gamma_{c}\psi+e_t ^{a}\zeta_a \,, \nonumber \\
\bar{\Lambda} & = & v\bar{\psi}-\frac{i}{2}A_t \bar{\psi}+\frac{1}{2}\omega_t ^{c}\bar{\psi}\Gamma_{c}+\bar{\zeta_a}e_t ^{a} \, , \\
\Lambda_{j}^{b} & = & D_{j}\omega_t ^{b}+2e_t ^{c}e_{j}^{a}\epsilon_{\:ca}^{b}\bar{\psi}\psi \,, \nonumber\\
\lambda_{j} & = & \partial_{j}A_t +2ie_t ^{a}e_{j}^{b}\epsilon_{abc}\bar{\psi}\Gamma^{c}\psi \,. \nonumber
\end{eqnarray}

After inserting \eqref{fixmultipliers} into the consistency conditions \eqref{consistency}, these reduce to

\begin{eqnarray}
 0 &= & e_t ^{b}e_{j}^{c}(\eta_{ad}T_{\,cb}^{d}\bar{\psi}\psi-\bar{\psi}\Gamma_{a}\Gamma_{[c}\zeta_{b]}-\bar\zeta_{[b}\Gamma_{c]}\Gamma_{a}\psi) \,, \label{reduced1}\\
0 & = &  |e| \epsilon^{abc}(\Gamma_{ab}\zeta_c - \frac{1}{2}\Gamma_{a}\Gamma_{d}T_{\,bc}^{d}\psi) \,. \label{reduced2}
\end{eqnarray}
together with the conjugate of the last equation. For an arbitrary dreibein, equation \eqref{reduced2} can be used to fix the remaining free component of $\zeta_a$ as a function of the dynamical fields and $T^a_{\,\,\,bc}$. On the other hand, using the constraint $K_a\approx 0$, one can show that \eqref{reduced1} correspond to $6$ independent equations for an equal number of free components in $T^a_{\,\,\,bc}$, once $\zeta_a$ is replaced.\\

Note now that the parameter $v$ does not show up in (\ref{reduced1},\ref{reduced2}), this indicates that the complete set of equations is not independent. In fact, one can readily check that the following shift
\begin{eqnarray}\label{weylsym}
T^a_{\,\,\,bc} \rightarrow  T^a_{\,\,\,bc} + 2\beta \delta^a_{[b} E_{c]}^t &, \quad & \zeta_c \rightarrow \zeta_c - \beta E_c^t \psi \,,
\end{eqnarray}
leaves (\ref{ansatzT},\ref{ansatzD}) and (\ref{reduced1},\ref{reduced2}) invariant. This is related to the Weyl invariance, shifting the multiplier $v \rightarrow v - \beta$ in \eqref{fixmultipliers}. We thus have the following picture: If the three components $e^a_t$ remain arbitrary, then one can solve the $(14+4n)$ multipliers as in \eqref{fixmultipliers}, but this leaves a degeneracy in $v$ to be fixed afterwards.  Otherwise one may restrict one of the components $e^a_t$ while leaving the scaling parameter $v$ completely free, as we explain below. In view of the counting argument of Section \ref{degree_freedom_section}, we expect in general that one combination among the 8 parameters $(A_t, \omega^a_t,e^a_t, v)$ will be found fixed in a generic sector, so that the number of functionally independent first-class constraints is reduced to $F=7$. Note that the degeneracy in $v$ also suggests there could be certain configurations of the dynamical fields such that (\ref{reduced1},\ref{reduced2}) have no solution: The consistency equations would lead to secondary constraints in this sectors.\\

The above reasoning is illustrated with the spin-$1/2$ sector described in Section \ref{half_spin_sector}. In that case one chooses the gauge $v=0$ a priori, and then proceed to count the DOF considering the residual symmetries. This gauge fixing is equivalent to choose the solution

\begin{eqnarray}
T^a_{\,\,\,bc} &=& 2\alpha \epsilon^a_{\,\,\,bc} \, ,  \\
\zeta_a &=& \Gamma_a \psi \, ,
\end{eqnarray}
for (\ref{ansatzT},\ref{ansatzD}) and (\ref{reduced1},\ref{reduced2}), provided (\ref{T=ee},\ref{D-psi}). Inserting this into the multipliers \eqref{fixmultipliers} and then into the total Hamiltonian \eqref{hamiltonian}, one directly gets the form \eqref{total_hamiltonian_sector}. Note that this Hamiltonian preserves the gauge, and possesses only 7 free parameters $(A_t, \omega^a_t, e^a_t)$ corresponding to the generators of residual symmetries.

On the other hand, in a generic sector one can always use the "degenerate gauge" (\ref{ea=xiei}) for counting purposes. Using $K_a \approx 0$, this election is readily seen to close the consistencies (\ref{reduced1},\ref{reduced2}) and puts the total Hamiltonian in the form \eqref{Ht1}. However, by doing so one needs to assume there is in fact a solution for $\zeta_a$ and $T^a_{\,\,\,bc}$, in order to extend the sector for non-degenerate choices with $|e|\neq 0$. Thus, in any generic sector a realization of the first-class constraints can be easily obtained by means of the degenerate gauge, leaving also $F=7$ free parameters $(A_t, \omega^a_t,\xi^i,v)$.

\section{Asymptotic charges on the bosonic vacuum}\label{apcharges}      
In order to compute the charges in the smeared generator \eqref{smear}, we demand that the boundary terms induced by its variations with respect to the dynamical fields be well defined. That is, the variations of $Q$ should compensate the boundary contributions coming from the integration by parts on the bulk term. Varying the action \eqref{smear} with (\ref{DJa}-\ref{DHi}), and upon integrating by parts one finds
\begin{eqnarray}\label{var_asymptotic}
\delta Q_{G}[\xi^i] &=& - \int_{\partial\Sigma}dx^l \xi^{i}[2(\bar{\psi}\Gamma_{il}\delta\psi+\delta\bar{\psi}\Gamma_{li}\psi)+\epsilon_{jl}\delta(e_{i}^{a}p_{a}^{j})-\epsilon_{il}\delta e_{k}^{a}p_{a}^{k}] \, , \label{deltaQG} \\
\delta Q_{\tilde{J}} [\lambda^a]&=& -2\int_{\partial\Sigma}dx^l\eta_{ab}\lambda^{a}\delta\omega_{l}^{b} \, , \\
\delta Q_{\tilde{K}} [\lambda]&=&- 2\int_{\partial\Sigma}dx^l \lambda \delta A_{l} \, ,
\end{eqnarray}
where $\partial\Sigma$ the boundary of the spatial slice $t=constant$. It is clear that $\delta Q_{\tilde{J}}$ and $\delta Q_{\tilde{K}}$ can be readily integrated, i.e. the $\delta$ can be removed, but for $\delta Q_{G}[\xi^i]$ we need to give certain boundary conditions. If we impose $\psi$ and $\bar \psi$ to vanish at the boundary\footnote{More generally, one could consider for instance $\psi \sim \bar \psi \sim O(\frac{1}{r^2})$, where the asymptotic region is defined by $r \rightarrow \infty$. Then the leading order in $p^i_j \approx 2\epsilon^{ij}\eta_{ab}e^b_j \bar \psi \psi $ depends on the fall-off of the dreibein, and if $e^a_i \sim O(r)$, all the asymptotic contributions in \eqref{deltaQG} still vanish.}, then we are led to also fix $p^k_a = 0$ by consistency with $\varphi^k_a \approx 0$. These (gauge-consistent) conditions then annihilate the charge associated to $G_i$, while leaving the variation of the dreibein completely undetermined, as was discussed for the vacuum configuration in the bulk region. The conditions can thus be regarded as the natural asymptotic extension of such sector. \\

With these global charges, the smeared generator $S$ has well defined functional derivatives and consistently acts on the fields $(A_i, \omega^a_i, \psi, \bar \psi, e^a_i, p^j_b)$ through the Dirac bracket. Upon gauge fixing the bulk term in $S$ is identically dropped, leading to the so-called asymptotic charge algebra
\begin{eqnarray}\label{asymptU1}
\{ Q_{\tilde{K}}[\lambda],  Q_{\tilde{K}}[\zeta] \}_D &=&\{ Q_{\tilde{K}}[\lambda],  S[0,\zeta;0] \}_D \nonumber \\
 &=& \delta_{\tilde{K}[\zeta]} Q_{\tilde{K}}[\lambda] \nonumber \\
&=&  2\int_{d\Sigma}dx^l \lambda \partial_l \zeta \nonumber \\
&=& C_{\tilde{K}}[\lambda,\zeta] \, ,
\end{eqnarray}
and similarly,
\begin{eqnarray}\label{asymptSOU1}
\{ Q_{\tilde{J}}[\lambda^a],  Q_{\tilde{J}}[\zeta^a] \}_D &=& Q_{\tilde{J}}[\epsilon^a_{\,\, bc}\lambda^b \zeta^c]+ 2\int_{d\Sigma}dx^l \eta_{ab} \lambda^a \partial_l \zeta^b \, , \\
\{ Q_{\tilde{J}}[\lambda^a],  Q_{\tilde{K}}[\lambda] \}_D &=& 0 \,.
\end{eqnarray}
We see that the asymptotic algebra precisely corresponds to that of the CS theory for the direct product $SO(2,1) \times U(1)$, including the central extensions.

A short remark regarding the gauge consistency of the boundary conditions is now appropriate. Obviously, the fermion cannot be excited in the asymptotic region by an $SO(2,1) \times U(1)$ transformation (nor by diffeomorphisms or Weyl scaling, in contrast with SUSY). Also, as $\tilde{J}_a$ and $\tilde{K}$ are first class they preserve the constraint $\varphi^i_a = 0$ and thus our boundary conditions are gauge invariant. As a consequence, we see that $\delta_{\tilde{K}} Q_G = \delta_{\tilde{J}} Q_G= 0$ is consistent, i.e., valid in any gauge. Now, one could ask if $\delta_G Q_{\tilde{K}} = \{ Q_{\tilde{K}},Q_{G}\}=0$ also holds, and the same for $\delta_G Q_{\tilde{J}}$. In order to see this, let us compute explicitly the transformations generated by the improved diffeomorphisms. They read\footnote{Note that the improved diffeomorphism corresponds to a covariant generalization of the Lie derivative.}
\begin{eqnarray}\label{improved}
\{ e^a_i, S[\xi^i;0] \}_D = D_i (\xi^j e^a_j) + \xi^j T_{ji} \,, \quad && \{ \psi , S[\xi^i;0] \}_D = \xi^i D_i\psi \,, \quad \{ \bar{\psi} , S[\xi^i;0] \}_D = \xi^i \bar{\psi} \overleftarrow{D}_i \, , \label{D1} \\
\{ A_i , S[\xi^i;0] \}_D = \xi^j F_{ji}  \,, && \quad  \{\omega^a_i, S[\xi^i;0] \}_D = \xi^j R^a_{ji} \label{Domega} \, .
\end{eqnarray}
By virtue of \eqref{Domega} and the constraints (\ref{Ja},\ref{K}), we conclude the $SO(2,1) \times U(1)$ gauge fields are invariant under improved diffeomorphisms in the asymptotic region. This in turn yields $\delta_G Q_{\tilde{K}} = \delta_G Q_{\tilde{J}}=0$, as expected. A far simpler argument can be repeated for the Weyl generator $\Upsilon$, which is consistent with the algebra \eqref{simplealg}.

\end{document}